\documentstyle[sprocl]{article}

\def\Journal#1#2#3#4{{#1} {\bf #2}, #3 (#4)}

\def\NPB{{\em Nucl. Phys.} B}
\def\PLB{{\em Phys. Lett.}  B}
\def\PRL{\em Phys. Rev. Lett.}
\def\PRD{{\em Phys. Rev.} D}

\def\be{\begin{equation}}
\def\ee{\end{equation}}
\def\bea{\begin{eqnarray}}
\def\eea{\end{eqnarray}}

\font\mybb=msbm10 at 10pt
\def\bb#1{\hbox{\mybb#1}}
\def\bR {\bb{R}}
\def\bE {\bb{E}}

\newcommand{\nn}{\nonumber \\}

\begin{document}

\title{KILLING SPINORS, SUPERSYMMETRIES AND ROTATING INTERSECTING BRANES}

\author{P.K. TOWNSEND}

\address{DAMTP, University of Cambridge,
\\ Silver Street, Cambridge CB3 9EW, UK
\\E-mail: pkt10@damtp.cam.ac.uk}


\maketitle\abstracts{I review a recently proposed 
method for determining the
symmetry superalgebra of a supergravity configuration  
from its Killing spinors,
and its application to  the `near-horizon' limits of various rotating and
intersecting branes.}

A configuration of gravitational and matter fields is said to possess a
symmetry if there is an infinitesimal coordinate 
transformation, associated with
some non-zero vector field $\xi$, that leaves invariant all fields, including
the metric. The invariance of the metric implies that $\xi$ is Killing and
the set of all such fields closes under commutation to define the symmetry
algebra of the field configuration. For a supergravity theory one might
analogously define a `supersymmetric' configuration to be one for which there
is an infinitesimal local supersymmetry transformation, asociated with a 
non-zero anticommuting spinor field $\epsilon$, that leaves invariant all
fields. In this case, however, it is far from clear how one extracts the
symmetry superalgebra from the knowledge of these spinors. To begin with,
there is no way to construct an ordinary vector field from anticommuting 
spinors. To circumvent this problem we may note that $\epsilon$ is defined by a
set of {\it linear} differential equations, so a supersymmetric field
configuration is naturally associated with a common zero mode of a set  of
differential operators. This zero mode is a {\it commuting} spinor, but it
will generally have both a `body' and a `soul'. The body is an `ordinary'
commuting spinor, $\zeta$, satisfying the same equations as $\epsilon$ but
with all background fermions set to zero. For such purely bosonic backgrounds
the supersymmetry variations of the bosons vanish identically so the equations
determining $\zeta$ are found from the supersymmetry variations of the
fermions. These are the Killing spinor equations, and the solutions are
the Killing spinors. Thus, the number of supersymmetries of a purely
bosonic background equals the dimension of the space of linearly independent
Killing spinors.

Killing vectors may now be constructed as bilinears of Killing spinors. It is
not difficult to verify that the vector field $\bar\zeta\Gamma^m\zeta'
\partial_m$ is Killing if the spinor fields $\zeta$ and $\zeta'$ are. This
suggests that  the symmetry superalgebra is determined by the Killing spinor
fields (up to purely bosonic factors) just as the bosonic symmetry algebra is
determined by the Killing vector fields. One way to see how 
this works \cite{GMTa} starts by considering the supergravity field 
configuration of interest as a background for small
fluctuations. Given a Killing spinor or vector field of the background it is
then possible to construct a time-independent charge as a functional of the
fluctuation fields. One thus finds a set of fermonic charges $Q_F(\zeta)$ and a
set of bosonic charges $Q_B(\xi)$ for which the (anti)commutators may be 
computed via the canonical (anti)commutation relations of the fluctuation
fields \cite{GMTa}. For the bosonic charges one has
\be
[Q_B(\xi),Q_B(\xi')] = Q_B([\xi,\xi'])\, ,
\ee
so that the algebra of bosonic charges is homomorphic to the algebra of Killing
vector fields. For the mixed bosonic, fermionic commutator one has
\be
[Q_B(\xi),Q_F(\zeta)] = Q_B({\cal L}_\xi \zeta)\, ,
\ee
where ${\cal L}_\xi$ is the Lie derivative with respect to $\xi$. Note that the
Lie-derivative of a spinor field with respect to a vector field is defined
only if latter is Killing, as is the case here. Finally, and most
importantly, one has
\be
\{Q_F(\zeta),Q_B(\zeta')\} = Q_B(\xi)
\ee
where
\be
\xi^m = \bar\zeta\Gamma^m\zeta'\, .
\ee
What this shows \cite{GMTa} is that {\it the linear 
combination $\xi$ of Killing vector
fields associated to any pair of Killing spinor fields, $\zeta$ 
and $\zeta'$, is
the same as the linear combination $Q_B(\xi)$ of 
bosonic charges appearing in the
anticommutator of the fermionic charges $Q_F(\zeta)$ and $Q_F(\zeta')$}. In
this way, the symmetry superalgebra may indeed be deduced, up to purely bosonic
factors, from the Killing spinors. 

In many cases this fact is not needed because a knowledge of the bosonic
symmetry algebra and the number of supersymmetries suffices to determine
the superalgebra. A simple example is
the maximally supersymmetric $adS_2\times S^2$ Bertotti-Robinson 
solution of N=2
D=4 supergravity, which arises as the near-horizon limit of the extreme
Reissner-Nordstr{\"o}m (RN) black hole \cite{Carter,gib}. This information is
sufficient to determine the symmetry superalgebra; it can only be
$su(1,1|2)$, for which the bosonic subalgebra is $su(1,1)\oplus
su(2)$. 
Many higher-dimensional maximally supersymmetric $adS_{p+2}$
spacetimes arise in a similar way as near-horizon limits of p-brane 
solutions of
supergravity theories \cite{GT}, and if one runs through the list of anti-de
Sitter superalgebras \cite{Nahm} then one usually finds that there is only one
candidate. An interesting exception occurs for certain $adS_2$ and $adS_3$
cases. As $adS_3$ superalgebras are direct sums of two $adS_2$
superalgebras, we
may concentrate on the $adS_2$ case. There is a one-parameter family of $adS_2$
superalgebras,
\be
D(2|1;\alpha) \supset su(1,1) \oplus su(2) \oplus su(2)\, .
\ee
The parameter $0<\alpha\le 1$ is the relative weight of the the two $su(2)$
subalgebras. When $\alpha=1$ we have the isomorphism
\be
D(2|1;\alpha) \cong osp(4|2;\bR)\, ,
\ee
and it is convenient to define
\be
D(2|1;0) = su(1,1|2)\oplus su(2)\, .
\ee
Clearly, the parameter $\alpha$ is not determined by a knowedge of the bosonic
symmetry algebra and the number of supersymmetries. The simplest case in which
this ambiguity arises is the D=5 extreme Tangherlini black hole, which is the
D=5 generalization of the extreme RN black hole. This is a solution of D=5
supergravity with a maximally supersymmetric $adS_2\times S^3$ near-horizon
limit \cite{cham}. This near-horizon solution therefore has eight
supersymmetries and is invariant under the isometry group of $adS_2\times S^3$.
The superalgebra must be $D(2|1;\alpha)$ for some $\alpha$, but what
is the value of $\alpha$?

In this case there is an indirect argument that $\alpha$ must vanish. This
comes from consideration of angular momentum. It will prove useful to first
consider rotation in the context of D=4 black holes. The general stationary D=4
black hole solution depends on one angular momentum parameter, in addition to
its mass and charge parameters, but the requirement of supersymmetry forces the
angular momentum to vanish. This could seem puzzling in view of the fact that
angular momentum does not appear in the anticommutator of supersymmetry charges
of the super-Poincar{\'e} algebra, because this fact implies that every
non-rotating supersymmetric solution is a member of a class of rotating
supersymmetric solutions. Indeed it is, but all the rotating solutions have
naked singularities and so fail to qualify as `black holes'. As an aside, I
remark that angular momentum {\it does} appear in the anticommutator of
supersymmetry charges of the adS superalgebra, so in this case the value of the
angular momentum is of direct relevance to supersymmetry. In fact, it 
turns out \cite{KP}
that asymptotically adS supersymmetric D=4 black holes are {\it necessarily}
rotating; now it is the non-rotating
supersymmetric solutions that are singular! 

Returning to D=5, we first note that the general stationary asymptotically
flat solution will depend on two angular momenta \cite{MP},
corresponding to rotation in two independent 2-planes. When both vanish we
have a static black hole with $SO(4)$ hyperspherical symmetry. When they are
non-zero but equal the symmetry is reduced to $SO(3)\times SO(2)$ and when they
are unequal it is further reduced to $SO(2)\times SO(2)$. In the context of D=5
supergravity these groups become
\be
SU(2)\times SU(2) \supset SU(2)\times U(1) \supset U(1)\times U(1)\, .
\ee
It turns out that supersymmetry requires both 
angular momenta to be equal \cite{Rob}. Let $j$ 
be the one independent angular momentum parameter of the
supersymmetric black hole solutions; for an appropriate normalization one finds
that non-singularity on and outside the event horizon requires $|j|<1$. When
$j=0$ we have the Tangherlini BH with its $adS_2\times S^3$ 
near-horizon geometry.
When $j\ne0$ we have a rotating version with a near horizon geometry that is 
no longer a direct product. Let $\sigma_i$ ($i=1,2,3$) be the three
left-invariant 1-forms on $S^3$ satisfying $d\sigma_1 = \sigma_2\wedge
\sigma_3$ and cyclic permutations; then $d\Omega_3^2 = (1/4)(\sigma_1^2
+\sigma_2^2 + \sigma_3^2)$ is the $SU(2)\times SU(2)$ invariant metric on the
unit sphere. In this notation the near-horizon metric of the rotating
supersymmetric D=5 black hole is \cite{GMTb}
\be\label{five}
ds^2 = -(r^2 dt + {j\over2}\sigma_3)^2 + r^{-2}dr^2 + d\Omega_3^2
\ee

This metric is clearly $SU(2)\times U(1)$ invariant, and at least 1/2
supersymmetric, but its full isometry supergroup is far from evident.  An
analysis of the Killing spinor equations reveals that supersymmetry is
again fully
restored in the near-horizon limit \cite{kal}. This is
remarkable because it suggests that rotation reduces the near-horizon isometry
group without reducing its supersymmetry. This would be hard to understand
unless the $SU(2)$ factor that is broken to $U(1)$ by the rotation is a purely
bosonic factor of the isometry supergroup, and this is the case only if
$\alpha=0$. In fact, a direct determination \cite{GMTb} 
of the symmetry superalgebra using
the method described above confirms that the near-horizon 
symmetry superalgebra of
the D=5 supersymmetric black hole is $su(1,1|2)\oplus su(2)$, when $j=0$, and 
$su(1,1|2)\oplus u(1)$ when $j\ne0$. It also determines the $su(1,1)$ Killing
vector fields to be\footnote{The notation used here differs from that
of ref. 10.}
\bea
h &=& \partial_t \nn
d &=& r\partial_r -2t\partial_t \nn
k &=& [(1-j^2)r^{-4}+ 4t^2]\partial_t -4tr\partial_r + 2jr^{-2}m\, ,
\eea
where $m$ is the conventionally normalized Killing vector field
associated with the angular momentum parameter $j$. The remaining 
$su(2)$ Killing vector fields in the bosonic subalgebra of $su(1,1|2)$
are the right-invariant vector fields on $S^3$. 

Supersymmetric D=5 black holes have various interpretations as intersecting
M-branes. The most straightforward 
is as three intersecting \cite{pap}, and possibly 
rotating \cite{cvetic}, M2-branes, customarily
denoted by $(0|M2,M2,M2)$. Compactification on the 6-dimensional space
spanned by the M2-branes yields a 1/8 
supersymmetric solution of $T^6$-compactified 
D=11 supergravity, and a subclass (the `G{\"u}ven' class)
yields 1/2 supersymmetric solutions of the equations of pure D=5
supergravity. An instructive, although less direct, route begins from
the dual $(1|MW,M2,M5)$ configuration. Wrapping the M2-brane on a 
circle and the M5-brane on a 4-torus yields a 1/8 supersymmetric
string solution of $T^5$-compactified D=11 supergravity. The
`G{\"u}ven' subclass now yields a 1/2 supersymmetric solution of the 
pure (1,1) supersymmetric D=6
supergravity for which the near-horizon geometry is $adS_3\times S^3$,
irrespective of the rotation \cite{larsen} because its effects can 
be removed {\it in the
near-horizon limit} by a coordinate transformation. The symmetry
superalgebra of this near-horizon solution is 
\be\label{superisom}
su(1,1|2)_L \oplus su(1,1|2)_R\, .
\ee
The total number of supersymmetries is thus 16, which is double that
of the D=5 black hole, but
only the `left' supersymmetries survive further reduction on $S^1$. Moreover,
the identification needed to perform this reduction is not preserved by the
coordinate transformation needed to remove the effects of rotation in the
near-horizon limit \cite{GMTa}, so that the near-horizon geometry in 
D=5 depends on the rotation parameter, in agreement with the 
direct analysis in D=5. In fact, 
$SU(1,1|2)_R$ is broken to $SU(2)$ for zero rotation 
and to $U(1)$ for non-zero
rotation. We thus recover the  D=5 results deduced earlier. 

The 1/8 supersymmetric $(1|MW,M2,M5)$ 
configuration of M-theory is a special case
of the 1/8 supersymmetric $(1|MW,M2,M5,M5)$ configuration. For an appropriate
choice of the various (generalized) harmonic functions associated with the
M-Wave, M2-brane and M5-brane components, there is a non-singular near-horizon
limit with $adS_3\times S^3\times S^3\times\bE^2$ 
geometry \cite{CT,BPS}, with the
ratio of the two $S^3$ radii equal to the ratio of the two M5-brane
charges. The number of independent Killing spinors is 16, 
which represents a {\it fourfold} increase in
supersymmetry relative to the full intersecting brane configuration
(as was the case for the $(1|MW,M2,M5)$ configuration). The reason for this
is that the M-Wave is irrelevant in the near-horizon limit because 
there are no local degrees of freedom of the gravitational field in
$adS_3$. 

The 1/2 supersymmetric  $adS\times S^3\times S^3\times\times \bE^2$ 
solution of D=11 supergravity can obviously be viewed as a 1/2
supersymmetric $adS_3\times S^3\times S^3$ solution of D=9 maximal 
supergravity. We now face the problem of determining the isometry
superalgebra of this solution. Indirect arguments \cite{BPS} suggest 
that it should be
\be
D(2|1;\alpha)\oplus D(2|1;\alpha)
\ee
with $\alpha$ the ratio of the two $S^3$ radii. The removal of one
M5-brane corresponds to taking this ratio to zero, and hence to $\alpha=0$,
which we earlier saw to be  correct. However, these
arguments are only suggestive and this example provides what is perhaps the
best illustration of the need for a systematic method. An application of 
the method described earlier confirms \cite{GMTa} that the 
superalgebra is indeed $D(2|1;\alpha)$, with $\alpha$ the ratio 
of the two $S^3$ radii. Thus all values of the parameter $\alpha$ of
the $adS$ superalgebra $D(2|1;\alpha)$ have a physical realization 
within M-theory.

\end{document}